\documentclass[preprint,showkeys,secnumarabic,amsfonts,amsmath,amssymb]{revtex4}
\usepackage[dvips]{color}
\usepackage{array}
\usepackage{epsfig}
\usepackage{amsmath}
\usepackage{graphicx}
\begin{document}
\title{Dynamical System Analysis of Randall-Sundrum Model with Tachyon Field on the Brane}

\author{A. Ravanpak}
\email{a.ravanpak@vru.ac.ir}
\affiliation{Department of Physics, Vali-e-Asr University of Rafsanjan, Rafsanjan, Iran}
\author{G. F. Fadakar}
\email{g.farpour@vru.ac.ir}
\affiliation{Department of Physics, Vali-e-Asr University of Rafsanjan, Rafsanjan, Iran}

\date{\today}

\begin{abstract}

In this manuscript we use the dynamical system approach to study the linear dynamics of a Randall-Sundrum braneworld model with a tachyon scalar field confined to the brane. We recognize that the form of the tachyon potential plays a significant role in the evolution of the universe. For the case of an inverse square potential we find that one of our new variables, $\lambda$, is constant. We obtain critical points of the system in this situation and investigate their stability using the linear perturbation method. Then we turn to a Gaussian potential in which $\lambda$ is not constant. Using the idea of instantaneous critical points we study the behavior of the universe and its possible fates. One of the interesting results of this manuscript is that our universe will probably experience another phase transition from acceleration to deceleration in the future.

\end{abstract}

%\pacs{98.80.-k; 45.30.+s; 95.36.+x}

\maketitle

\section{Introduction}

Nowadays the big bang theory is the predominant cosmological model according to which our universe began to expand from a singularity with an extreme temperature and density. It left behind a very rapidly accelerated expansion era called inflation, a radiation dominated phase and a matter dominated era one after another till it reached the current dark energy (DE) dominated phase in which it is experiencing another accelerating expansion. There are many candidates for driving the evolution of the universe specially in its accelerated expanding phases that have been interviewed and investigated in the literature in detail \cite{Parsons}-\cite{Gao}. Among them, the idea of a tachyon scalar field, $\phi$, is of particular interest, because it can be considered as an inflaton field to produce an inflationary era and also can play the role of the dark matter and the DE of the universe \cite{Gibbons}-\cite{Ravanpak}. The tachyon field that has its roots in string theory possesses special properties. The tachyonic potential, $V(\phi)$, is always positive with a maximum at $\phi=0$, and has a zero value when $\phi\rightarrow\infty$. Also, the derivative of the potential with respect to $\phi$, $V_\phi$, is always negative.

Besides, string theory has another substantial application in the field of cosmology and it is the idea of higher dimensional gravity and brane cosmology. Although Kaluza and Klein, first brought up the concept of an extra dimension in their famous theory, it was developed after the appearance of the string theory in the attempts of Arkani-Hamed, Dimopoulos and Dvali \cite{Arkani}, and more importantly in the works of Randall and Sundrum (RS), who proposed two useful 5D cosmological models \cite{Randall}-\cite{Randall2}. In their second model (RSII), which is the case of interest in this manuscript our universe is assumed to be a 4D brane in an infinite 5D spacetime, called bulk. The standard model of particle physics is restricted to the brane and just gravitons can propagate into the bulk. The application of the brane scenario in explaining the evolution of the universe has been demonstrated in the literature \cite{Setare}-\cite{Saridakis2}.

Regardless of many articles in which in the context of either a common 4D scenario or a brane cosmology and in the presence of various candidates for driving accelerating expansion, the behavior of the universe has been studied in an inflationary era or a DE dominated regime separately, another useful mathematical method that recently has been widely used in cosmological research, is the dynamical system approach in which one can investigate the whole history of the universe, all at once \cite{Wainwright}-\cite{Chimento}. In this method we can find all the possible trajectories of the universe related to different initial conditions in an appropriate phase space to study its long term behavior from the beginning until now, and not just one trajectory such as the case in Newtonian mechanics. What is important is to distinguish the type of the trajectories and classify them, using stability analysis.

In \cite{Copeland}-\cite{Nozari}, the evolution of the universe in the presence of a tachyon scalar field as the DE component and/or the inflaton field have been studied in the context of the dynamical system approach. The dynamical system perspective of a self-interacting scalar field in a RSII braneworld model and also in another interesting 5D braneworld model called DGP, has been investigated in \cite{Gonzalez} and \cite{Quiros}, respectively. Specifically, in a recent article we have studied the details of a DGP braneworld cosmology with a tachyon field on the brane using this method in \cite{Ravanpak4}. In all of these articles there is at least one stable critical point that behaves as a late time attractor, in addition to some other saddle or unstable critical points that correspond to other cosmological periods.

Another interesting issue that may appear in dynamical system analysis (considering some assumptions) is the concept of instantaneous critical points. This happens when a critical point depends upon one of the dynamical variables. For a scalar field in both 4D and 5D scenarios, it is the form of the potential that identify such critical points. In \cite{Copeland}, the authors have investigated the presence of instantaneous critical points for various 4D scalar field DE models. The same work has been done in \cite{Copeland2}, for only a 4D tachyon scalar field DE scenario, but with different types of potentials. In \cite{Ravanpak4}, we have also studied the existence of instantaneous critical points in a tachyon DGP cosmology with a Gaussian potential.

Here, we will analyze the stability of a RSII braneworld model with a tachyon field on the brane. Specially, the case of a Gaussian tachyonic potential will be studied. In Sec.\ref{s2}, we will review the basic equations of the model. Sec.\ref{s3}, is the main part of this article. It consists of defining a suitable set of new variables, deriving an autonomous system of ordinary differential equations, and investigating the stability of the model in the respective phase space. We will divide our discussion into two parts. In \ref{s3s1}, an inverse square potential will be considered and the critical points of the model will be obtained. In \ref{s3s2}, we will continue with a Gaussian potential and considering the concept of instantaneous critical points, we will reach interesting results. Through the paper we use natural units ($8\pi G = \hbar = c =M_p$ =1).

\section{THE MODEL}\label{s2}

The Friedmann equation on the brane in the Randall-Sundrum scenario can be written as follows
\begin{equation}\label{fried}
3H^2=\rho_{tot}(1+\frac{\rho_{tot}}{2\sigma})
\end{equation}
Here $H$, is the Hubble parameter, $\sigma$ is the brane tension and $\rho_{tot}=\rho_m+\rho_{tac}$, is the total energy density on the brane in which $\rho_m$, is the matter energy density and $\rho_{tac}$, denotes the energy density of the tachyon field. In the absence of any interaction between the dark sectors of the universe they satisfy conservation equations separately as follows
\begin{eqnarray}
  \dot\rho_m &+& 3H\rho_m = 0 \label{conservationdm} \\
  \dot\rho_{tac} &+& 3H(\rho_{tac}+P_{tac})= 0 \label{conservationtac}
\end{eqnarray}
The pressure of the tachyon field, $P_{tac}$, and its energy density are given by
\begin{eqnarray}
  \rho_{tac} &=& \frac{V(\phi)}{\sqrt{1-\dot\phi^2}} \label{rho} \\
 P_{tac} &=& -V(\phi)\sqrt{1-\dot\phi^2} \label{p}
\end{eqnarray}
in which the dot means derivative with respect to the cosmic time.
Replacing Eqs.(\ref{rho}) and (\ref{p}), into Eq.(\ref{conservationtac}), one can obtain the equation of motion of the tachyon field as
\begin{equation}\label{field}
\frac{\ddot\phi}{1-\dot\phi^2}+3H\dot\phi+\frac{V_\phi}{V}=0
\end{equation}
Also, the Raychaudhury equation that is very useful in the following sections can be obtained as
\begin{equation}\label{ray}
\dot H = -(\frac{1}{2}+\frac{\rho_{tot}}{2\sigma})(\rho_m+\rho_{tac}\dot\phi^2)
\end{equation}

\section{phase space and Stability analysis}\label{s3}

As we mentioned in the introduction, our objective is to rewrite the defined system above as an autonomous system of ordinary differential equations to analyze the stability of our model in an appropriate phase space. So, we define the following new set of dimensionless dynamical variables
\begin{equation}\label{nv}
y=\frac{\sqrt{V}}{\sqrt3H}, \quad z=\frac{\rho_{tot}}{3H^2}, \quad d=\dot\phi, \quad \lambda=-\frac{V_\phi}{V^{3/2}}
\end{equation}
Using the Friedmann equation and the new variables above, one can obtain
\begin{equation}\label{c}
\frac{\rho_{tot}}{2\sigma} = \frac{1-z}{z}
\end{equation}
and find a constraint for $z$, as $0\leq z\leq1$. The case $z=1$, corresponds to the low-energy limit ($\rho_{tot}\ll\sigma$) and shows a standard 4D Einstein-Hilbert theory coupled to a tachyon field. On the other hand, the high-energy limit ($\rho_{tot}\gg\sigma$) relates to the situation $z=0$. We have to notice that $z=0$, is achieved when $H\rightarrow\infty$ (and not $\rho_{tot}=0$), which corresponds to the very early universe, the beginning of inflation or even earlier. Also, from the definitions of the new variables above when $z=0$ ($H\rightarrow\infty$), we have certainly $y=0$. Furthermore, the Friedmann constraint can be obtained by rewriting Eq.(\ref{fried}), in terms of the new variables as
\begin{equation}\label{fried-cons}
z-\frac{y^2}{\sqrt{1-d^2}}=\frac{\rho_m}{3H^2}
\end{equation}
The right hand side of this equation is $\Omega_m$, which satisfies the constraint $0\leq\Omega_m\leq1$. Moreover, $-1\leq d\leq1$, because of the square root. Combining these constraints and the one of $z$, and considering Eq.(\ref{fried-cons}), we find out that $-1\leq y\leq1$. One can consider the positive values of $y$, as an expanding universe and the negative values as a contracting one.

\begin{figure*}[h]
\centering
\includegraphics[width=7cm]{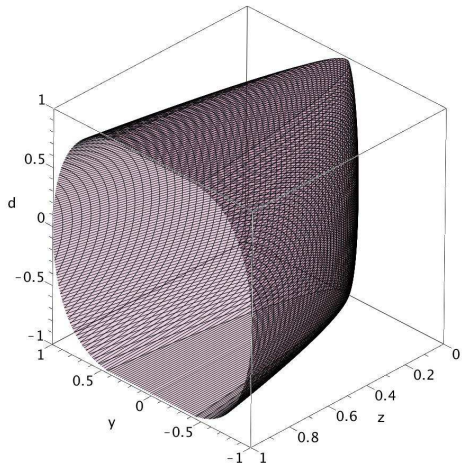}
\caption{The phase volume of our model}\label{fig1}
\end{figure*}

Regarding the Friedmann constraint and other constraints on our new variables $z$, $y$, $d$, and also the one of $\Omega_m$, we can identify our 3D phase space. FIG.\ref{fig1}, illustrates the respective phase volume. It is obvious that for $z=0$, we always have $y=0$, while $d$, can vary from $-1$ to $1$. Also, only for $d=0$ and $z=1$, $y$, can take the values $\pm1$.

Furthermore, we can obtain the tachyon equation of state (EoS) parameter and the total EoS parameter of the universe in terms of the new dimensionless variables as
\begin{eqnarray}
w_{tac} &=& d^2-1 \label{wtac}\\
w_{tot} &=& \frac{-y^2}{z}\sqrt{1-d^2} \label{wtot}
\end{eqnarray}
such that with attention to the constraints on the variables $y$ and $z$, we can constrain them as $-1\leq w_{tac}\leq0$, and $w_{tot}\leq0$. In \cite{Gumjudpai}, the author has evaluated the condition of acceleration for a few cosmological models such as a RSII braneworld model. If we rewrite this condition in terms of our phase space variables we obtain
\begin{equation}\label{acceleration}
    w_{tot}<-\frac{2}{3}+\frac{z}{3(2-z)}
\end{equation}
that guarantees an accelerated expansion. This condition for $z=1$, reduces to $w_{tot}<-1/3$, that is in agreement with the case of a standard 4D scenario. In addition to Eqs.(\ref{wtac}) and (\ref{wtot}), and with attention to Eq.(\ref{ray}), we can reach another useful relation which will be utilized in the following calculations as
\begin{equation}\label{hdot}
\frac{\dot H}{H^2} = \frac{3}{2}\left(\frac{z-2}{z}\right)(z-y^2\sqrt{1-d^2})
\end{equation}
Now, using Eqs.(\ref{fried-cons}) and (\ref{hdot}), we calculate a set of evolutionary equations for the model under consideration by differentiating the new variables in Eq.(\ref{nv}), with respect to $\ln a$. We find that
\begin{eqnarray}
  y' &=& -\frac{\sqrt{3}}{2}y^2d\lambda+\frac32y\left(\frac{2-z}{z}\right)(z-y^2\sqrt{1-d^2}) \label{yprime}\\
  z' &=& 3(1-z)(z-y^2\sqrt{1-d^2}) \label{zprime}\\
  d' &=& -(3d-\sqrt{3}y\lambda)(1-d^2) \label{dprime}\\
  \lambda'&=&-\sqrt3 dy\lambda^2(\Gamma-3/2) \label{lprime}
\end{eqnarray}
in which prime means derivative with respect to $\ln a$ ($a$, is the scale factor), $\Gamma=VV_{\phi\phi}/V_\phi^{2}$, and $V_{\phi\phi}$, represents the second derivative of the potential with respect to the tachyon field. These equations form a four dimensional autonomous system of ordinary differential equations and demonstrate the evolution of our phase space variables $d$, $y$, $z$ and $\lambda$, and so indirectly the behavior of the RSII model in the presence of a tachyon scalar field on the brane. But, because $z$, appears in the denominator of the second term in the right hand side of Eq.(\ref{yprime}), and since as we mentioned earlier it must satisfy the constraint $0\leq z\leq1$, Eq.(\ref{yprime}), will be undefined at $z=0$, and as a result we may miss possible critical points with $z=0$. So, we rewrite the system of equations above as
\begin{eqnarray}
  \tilde{y} &=& -\frac{\sqrt{3}}{2}y^2zd\lambda+\frac32y(2-z)(z-y^2\sqrt{1-d^2}) \label{yprime2}\\
  \tilde{z} &=& 3z(1-z)(z-y^2\sqrt{1-d^2}) \label{zprime2}\\
  \tilde{d} &=& -z(3d-\sqrt{3}y\lambda)(1-d^2) \label{dprime2}\\
  \tilde{\lambda} &=& -\sqrt3 dyz\lambda^2(\Gamma-3/2) \label{lprime2}
\end{eqnarray}
in which $\tilde{v}=zv', (v=y,z,d,\lambda)$.

The next step in the dynamical system approach is to find the critical points of the model and study their stability characteristics with attention to respective eigenvalues. Henceforth, it is more convenient to divide our discussion into two different situations. At the first stage, we investigate the case of a constant $\lambda$, and calculate related critical points of the model and study the evolution of the universe, for various values of the parameter $\lambda$. Then, we carry on with a varying $\lambda$ situation. In this case, we assume that $\lambda$, evolves slowly enough such that one can consider the previous critical points as the instantaneous critical points for the present case. Using the idea of moving critical points, we find interesting results and analyze them.

\subsection{The Constant $\lambda$}\label{s3s1}

\subsubsection{critical points}\label{cp}

When the tachyonic potential has an inverse square behavior, $V(\phi)=V_0\phi^{-2}$, $\lambda$ will be a nonzero constant parameter. One can easily check this by integrating the definition of $\lambda$, in Eq.(\ref{nv}). The case $\lambda=0$, relates to a constant potential that does not satisfy the general properties of a tachyonic potential. A constant $\lambda$, yields $\tilde{\lambda}=0$, so we can obtain the fixed points of the system by setting $\tilde{y}=\tilde{z}=\tilde{d}=0$, simultaneously. The results are as follows:

\begin{itemize}
  \item $P_1 (y=0,z=1,d=0)$ : This critical point corresponds to a matter dominated universe, because regarding the Friedmann constraint one can obtain $\Omega_m=1$, and also with attention to Eq.(\ref{wtot}), $w_{tot}=0$. The same result could be achieved when we utilize the definitions of our new variables in Eq.(\ref{nv}).
  \item $P_2^\pm (y=0,z=1,d=\pm1)$ : Although $w_{tot}=0$, but with attention to Eq.(\ref{nv}), the critical points $P_2^\pm$, relate to matter scaling solutions in which the energy density of the tachyon field mimics the matter energy density and can be characterized by $w_{tac}=w_m$. In $P_2^\pm$, $w_{tac}=0$, that is equal to the EoS parameter of the matter content in our model, $w_m=0$. Also, obviously it is the kinetic part of the tachyon field that contributes in these solutions.
  \item $P_3^\pm \left(y=\pm y_\ast,z=1,d=\pm d_\ast=\pm\frac{\lambda y_\ast}{\sqrt3}\right)$ $\left[y_\ast=\sqrt{\frac{\sqrt{\lambda^4+36}-\lambda^2}{6}}\right]$ : These critical points require more attention, because of the appearance of $\lambda$. It is a bit difficult but possible to prove that $y_\ast^2=\sqrt{1-d_\ast^2}$. Therefore, the Friedmann constraint for $P_3^\pm$, leads to $\Omega_m=0$, which means they are tachyon field dominated solutions. Whether the kinetic term or the potential term is dominant, directly depends on the value of $\lambda$, so that when $\lambda\rightarrow0$, we have $y_\ast\rightarrow1$ and $d_\ast\rightarrow0$, and as a result a tachyonic potential dominated solution, and for $\lambda\rightarrow\infty$, we find $y_\ast\rightarrow\sqrt3/\lambda\rightarrow0$ and $d_\ast\rightarrow1$, and consequently a kinetic dominated solution. Also, when we calculate the total EoS parameter for these critical points we reach $w_{tot}=-1+\lambda^2y_\ast^2/3$, which with attention to the general limits of $w_{tot}$ in our model, yields a new useful constraint as $0\leq\lambda^2y_\ast^2\leq3$. One can check that $\lambda=0$, leads to $\lambda^2y_\ast^2=0$; along with the increasing of $\lambda$, $\lambda^2y_\ast^2$ grows as well, and in the limit $\lambda\rightarrow\infty$, $\lambda^2y_\ast^2=3$. Although these critical points are always tachyon dominated, they demonstrate an accelerated expansion only for a specific range of the parameter $\lambda$. As it is clear from Eq.(\ref{acceleration}), the condition of acceleration for $P_3^\pm$, is $w_{tot}<-1/3$, that in turn using Eq.(\ref{wtot}), yields $\lambda<\sqrt{2\sqrt3}$.
\end{itemize}
Obviously, all these five critical points are associated with the standard 4D limit because of $z = 1$. Although we can not see the effect of an extra dimension in these fixed points directly, we can find its role indirectly. For instance, in a pure 4D scenario, both the matter dominated and the matter scaling solutions are repellers while in our model they behave as saddle points.

In addition, there are other critical points with $z=0$, that relate to the high energy regime and show the role of the extra dimension in our model clearly, and as we mentioned earlier in all of them $y=0$, because in the early universe $H\rightarrow\infty$. Also, $w_{tot}$, is undefined in all of them, so we discuss their physical interpretations using other cosmological parameters and the definitions of phase space variables, themselves. They are as follows:

\begin{itemize}
  \item $P_4 (y=0,z=0,d=0)$ : Substituting this critical point into the Friedmann constraint we obtain $\Omega_m=0$. Also, $d=0$ reveals that the kinetic term of the tachyon field does not contribute to this solution. Moreover, as we discussed above $y=0$, because of $H\rightarrow\infty$, and therefore the tachyonic potential is nonzero. So, one can consider $P_4$, as an inflationary solution, in which the tachyon field plays the role of inflaton field. This result is in agreement with $w_{tac}=-1$, for this critical point.
  \item $P_5^\pm (y=0,z=0,d=\pm1)$ : Although with attention to Eq.(\ref{fried-cons}), these values certainly do not yield $\Omega_m=0$, but we can consider it because of $H\rightarrow\infty$. It can also be deduced assuming that in the very early universe the contribution of the matter content is negligible. Thus, we can consider $P_5^\pm$, as tachyon field dominated solutions. Depending on the value of $V$, they might be kinetic dominated solutions. One can check that at these points, $w_{tac}=0$.
  \item $L_1 (y=0,z=0,d=d)$ : As it is clear, $L_1$, is a critical line. It leads to $\Omega_m=0$, so it can be considered as a tachyon scalar field dominated solution for which $-1\leq w_{tac}\leq0$. Obviously, $P_4$ and $P_5^\pm$, belong to $L_1$.
\end{itemize}

In fact all the points of our phase volume in the plane $z=0$, which is the line $(y=0,z=0,d)$, are critical points. Though the contributions of the kinetic and the potential parts of the tachyon field differ for these points, all of them are scalar field dominated solutions.

\subsubsection{stability around the critical points}\label{s}

To investigate the behavior of the system near the critical points obtained above we consider small linear perturbations $\delta y$, $\delta z$ and $\delta d$, around them. Then, using Eqs.(\ref{yprime}), (\ref{zprime}) and (\ref{dprime}), it is easy to obtain the differential equations for these perturbations as
\begin{equation}\label{perturbation}
    \frac{d}{d\ln a}\left(
                      \begin{array}{c}
                        \delta y \\
                        \delta z \\
                        \delta d \\
                      \end{array}
                    \right)=\left(
                              \begin{array}{ccc}
                                \frac{\partial\cal Y}{\partial y} & \frac{\partial\cal Y}{\partial z} & \frac{\partial\cal Y}{\partial d} \\
                                \frac{\partial\cal Z}{\partial y} & \frac{\partial\cal Z}{\partial z} & \frac{\partial\cal Z}{\partial d} \\
                                \frac{\partial\cal D}{\partial y} & \frac{\partial\cal D}{\partial z} & \frac{\partial\cal D}{\partial d} \\
                              \end{array}
                            \right)\left(
                      \begin{array}{c}
                        \delta y \\
                        \delta z \\
                        \delta d \\
                      \end{array}
                    \right)
\end{equation}
in which $\cal Y$, $\cal Z$ and $\cal D$, are the right hand side of Eqs.(\ref{yprime}), (\ref{zprime}) and (\ref{dprime}), respectively. The above $3\times3$ matrix is called the Jacobian matrix that has to be evaluated at the critical points. It possesses three eigenvalues $\mu_1$, $\mu_2$ and $\mu_3$, that appear in the general solutions for the evolution of $\delta y$, $\delta z$ and $\delta d$ as follows:
\begin{eqnarray}
  \delta y &=& c_1a^{\mu_1}+c_2a^{\mu_2}+c_3a^{\mu_3} \\
  \delta z &=& c_4a^{\mu_1}+c_5a^{\mu_2}+c_6a^{\mu_3} \\
  \delta d &=& c_7a^{\mu_1}+c_8a^{\mu_2}+c_9a^{\mu_3}
\end{eqnarray}
where $c_{i=1..9}$, are integration constants. The nature of the critical points depends on the sign of these eigenvalues so that if all of them have negative (positive) values, we have a stable (an unstable) critical point and if two of them have opposite signs, we have a saddle critical point. In addition, if at least one eigenvalue is zero (and the nonzero eigenvalues have the same signs), we can not investigate the stability properties of respective critical points using a linear perturbation method. In such a case, one has to adopt other approaches in stability analysis such as the centre manifold theory. Since, in this manuscript we find a few such critical points, and since other stability approaches are beyond the scope of this manuscript, we resort to the numerical results to recognize their stability status. In the following, after calculating the elements of the Jacobian matrix for our model, we find its eigenvalues for each of the critical points in the previous subsection as follows:
\begin{itemize}
  \item $P_1$ : The eigenvalues for this critical point are $\mu_1=3/2$, $\mu_2=-3$ and $\mu_3=-3$. Obviously, they are real and have opposite signs, so $P_1$, demonstrates a saddle point.
  \item $P_2^\pm$ : The eigenvalues related to these critical points are $\mu_1=6$, $\mu_2=-3$ and $\mu_3=3/2$. Clearly, $P_2^\pm$, behave as saddle points in our 3D phase space.
  \item $P_3^\pm$ : The case for these critical points is a bit complicated, because the related eigenvalues depend on $\lambda$. They are $\mu_1=-\lambda^2y_\ast^2$, $\mu_2=-3+\lambda^2y_\ast^2$ and $\mu_3=-3+\lambda^2y_\ast^2/2$. Using the constraint on $\lambda^2y_\ast^2$, that we obtained earlier, we can conclude that $P_3^\pm$, are stable critical points and behave as attractors.
\end{itemize}

\begin{figure*}[h]
\centering
\includegraphics[width=7cm]{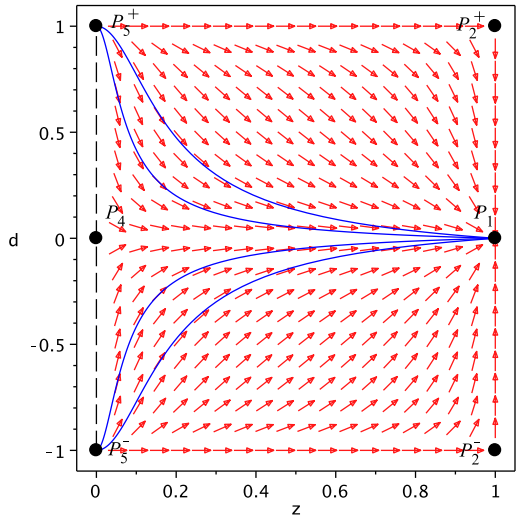}
\caption{The trajectories in the plane $y=0$, of the phase space. The dashed black line shows the critical subset $L_1$.}\label{fig2}
\end{figure*}

When we try to obtain the eigenvalues related to all the critical points in the plane $z=0$, i.e., $P_4$, $P_5^\pm$ and $L_1$, we reach $\mu_1=0$, $\mu_2=0$ and $\mu_3=0$. So, as we discussed above, the linear perturbation theory can not identify their stability characteristics. Instead, we resort to the numerical approach. All the trajectories in our phase space depend upon the value of $\lambda$, except the trajectories in two special planes $y=0$ and $d=0$, and fortunately we can recognize the stability status of all the critical points of the subset $(0,0,d)$, using just the trajectories in the plane $y=0$. FIG.\ref{fig2}, illustrates the evolution of trajectories in this plane. As it is clear, all the points on $L_1$, are saddle critical points, such as $P_4$, and only two critical points $P_5^\pm$, are repellers.

Finally, as we mentioned earlier, in a constant $\lambda$ scenario, $\lambda'=0$, which in turn as well as using Eq.(\ref{lprime}), yields $\Gamma=3/2$. This result is just as the one in \cite{Mizuno}, for a quintessence scalar field in the RSII model. But how about if we do not consider an inverse square potential?

\subsection{The Varying $\lambda$}\label{s3s2}

For any other form of tachyonic potentials except the inverse square, $\lambda$, is not a constant and evolves as other phase space variables. So, $\Gamma$, does not equal $3/2$ anymore and it may vary depending on the form of the potential. In the following we assume that $\lambda$ evolves sufficiently slow so that we can consider it as a constant in any infinitesimal period of time during the evolution of the universe. With this assumption all the critical points we obtained in the previous subsection, can be considered as instantaneous critical points for the present case. Among them, $P_3^\pm$, have dynamics because of $\lambda$ dependence. The concept of instantaneous and moving critical points helps us to understand how the universe tends to evolve at each instant.

Moreover, the type of the tachyon potential plays an important role in the evolution of our universe. For instance, in \cite{Bilic}, the authors have studied an inverse power law potential, $V(\phi)=V_0\phi^{-n}$ with $n>0$, in a tachyon braneworld cosmology and found a critical power $n_c$. They have demonstrated that for $0<n<n_c$, the asymptotic behavior of the universe is quasi de Sitter while for $n>n_c$, it is a dust universe. For a standard cosmology, $n_c=2$, as the authors have indicated in \cite{Abramo}. The same results have been obtained in \cite{Copeland2} in the context of dynamical system approach. Also, one can find different behaviors of the universe for some other types of potentials in \cite{Copeland2}. For example for an exponential potential $V(\phi)=V_0e^{-\mu\phi}$, the universe will eventually enter a non-accelerating regime.

Here, we choose a Gaussian potential, $V(\phi)=V_0\exp(-\alpha\phi^2)$, which satisfies all the tachyonic potential characteristics and in addition has a special property. It has an extremum at $\phi=0$, which relates to $\lambda=0$, the case we have ignored until now. When we evaluate the critical points of the model for the case $\lambda=0$, we find two additional critical points and also a critical line on top of the critical points we found earlier, as follows:
\begin{itemize}
  \item $P_6^\pm (y=\pm1,z=1,d=0)$ : Using the Friedmann constraint we conclude that at these critical points $\Omega_m=0$. On the other hand, $d=\dot\phi=0$. So, we find that $P_6^\pm$, are potential dominated or more exactly DE dominated solutions. It can be confirmed using Eqs.(\ref{wtac}) and (\ref{wtot}), because they lead to $w_{tac}=-1$ and $w_{tot}=-1$, that the latter shows an accelerated expansion. One can check that $P_3^\pm$, at the limit $\lambda\rightarrow0$, approach $P_6^\pm$.
  \item $L_2 (y=y,z=y^2,d=0)$ : As it is clear, $L_2$, is another critical line. Substituting $L_2$, into Eqs.(\ref{fried-cons}), (\ref{wtac}) and (\ref{wtot}), we again obtain $\Omega_m=0$, $w_{tac}=-1$ and $w_{tot}=-1$. Therefore, similar to $P_6^\pm$, $L_2$, is a DE dominated solution which corresponds to an accelerating universe, as well. What is important here is that the extra dimension, demonstrates its effect directly in this solution, such as $P_4$, $P_5^\pm$ and $L_1$. One can see that, $P_6^\pm$, are the end points of the critical line $L_2$, in the phase space.
\end{itemize}

\begin{figure*}[h]
\centering
\includegraphics[width=7cm]{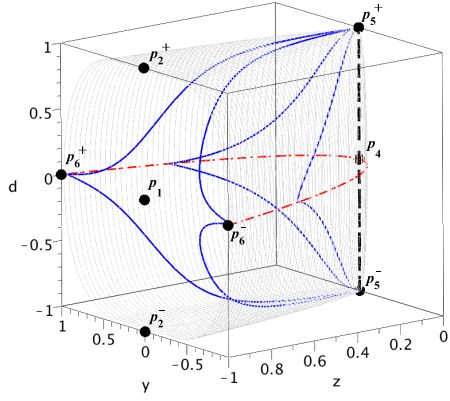}
\caption{The trajectories in the phase space of our model for the case $\lambda=0$. The dashed black line and the dash-dotted red curve show the critical lines $L_1$ and $L_2$, respectively.}\label{fig3}
\end{figure*}

When we evaluate the related eigenvalues for $P_6^\pm$ and $L_2$, we reach ($\mu_1=0$, $\mu_2=-3$, $\mu_3=-3$), and ($\mu_1=0$, $\mu_2=-3y^2$, $\mu_3=-3y^2$), respectively. Again, we resort to the numerical approach to distinguish the stability of $P_6^\pm$ and $L_2$. FIG.\ref{fig3}, which has been plotted for the special case $\lambda=0$, illustrates that all the possible trajectories of our universe start from the repellers $P_5^\pm$, and finally come to the critical line $z=y^2$ in the plane $d=0$. Therefore, $L_2$, and of course $P_6^\pm$, are attractor solutions.

To understand the evolution of the universe in our model completely, we need to know the asymptotic behavior of the parameter $\lambda$. In fact, its evolution for the case $\lambda\rightarrow0$, differs from the one of $\lambda\rightarrow\infty$. To find its asymptotic behavior we must refer to Eq.(\ref{lprime}). Using the definition of the Gaussian potential we calculate $\Gamma=1-1/(2\alpha\phi^2)$. So, $\Gamma-1$, and actually $\Gamma-3/2$ in Eq.(\ref{lprime}), are negative. Also, $\lambda^2$, is always positive. Now, if we consider an expanding universe for which $y>0$, and assuming $d>0$, we find that $\lambda'>0$, and therefore $\lambda$ approaches infinity, asymptotically. On the other hand, we can consider the case $\lambda=0$, as our starting point, because it relates to the top of the Gaussian potential. It can be considered as the preinflationary era or the beginning of inflation.

\begin{figure*}[h]
\centering
\includegraphics[width=7cm]{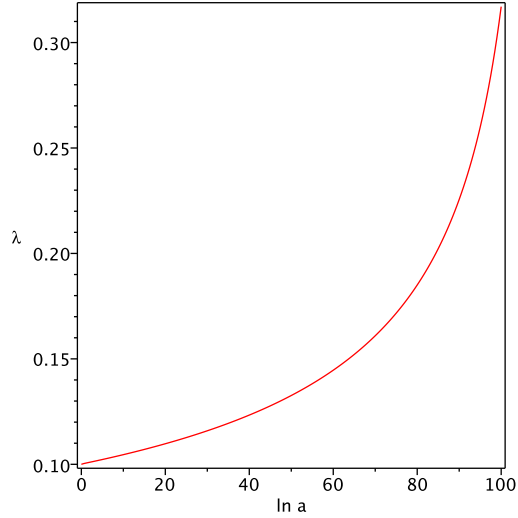}
\caption{The behavior of $\lambda$ in terms of $\ln a$. We have used the initial conditions: $y(0)=0.9$, $d(0)=0.1$, $z(0)=0.8$, $H(0)=70$ and $\lambda(0)=0.1$. Also, we have set $\alpha=1$.}\label{fig4}
\end{figure*}

Moreover, we can numerically demonstrate that considering a Gaussian potential is consistent with the assumption of a slowly varying $\lambda$ in the model under consideration. FIG.\ref{fig4}, illustrates the behavior of $\lambda$ versus $\ln a$. It is obvious that for a long range of $\ln a$, $\lambda$ changes very slowly.

\begin{figure*}
\centering
\includegraphics[width=7cm]{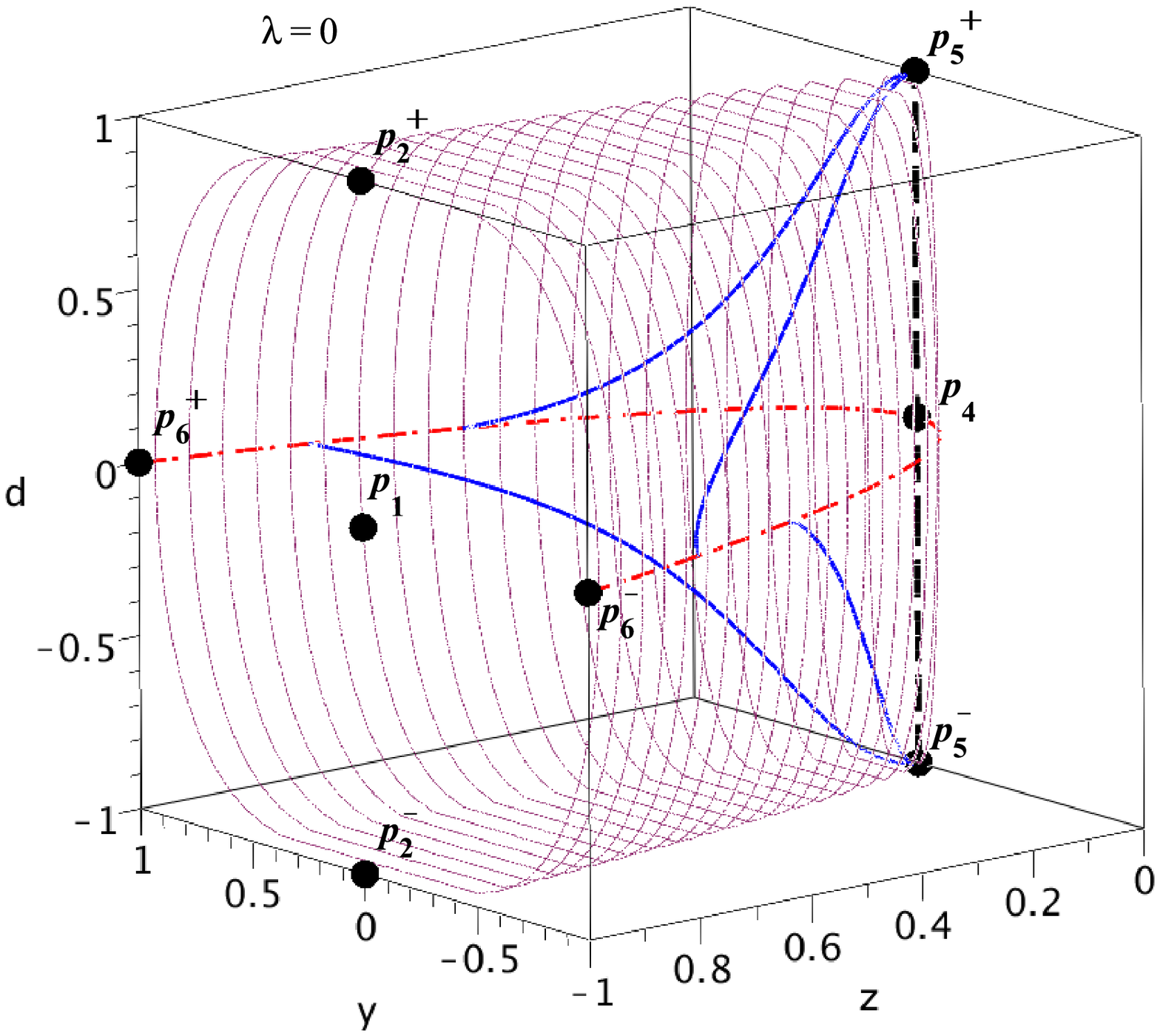}
\includegraphics[width=7cm]{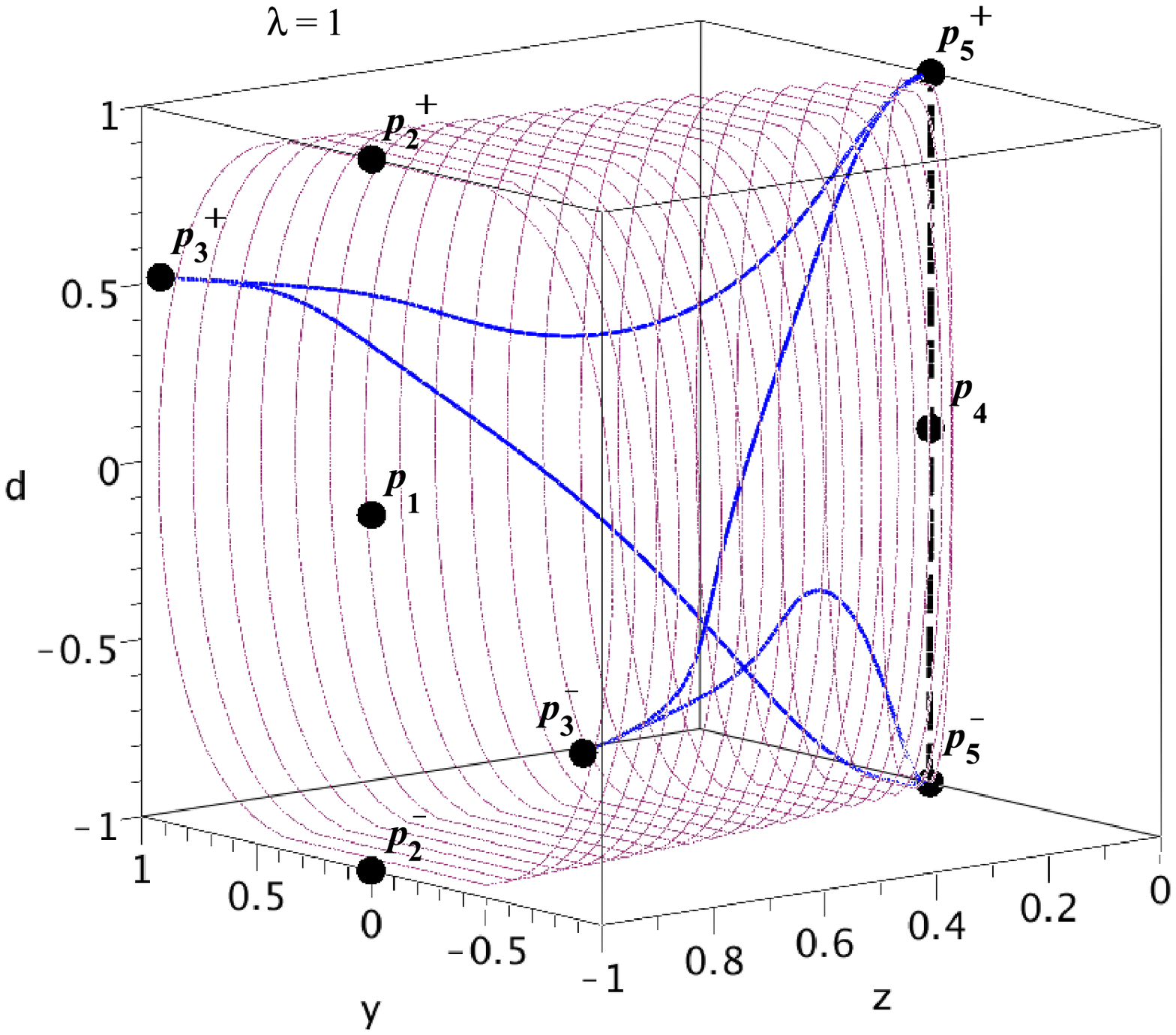}
\includegraphics[width=7cm]{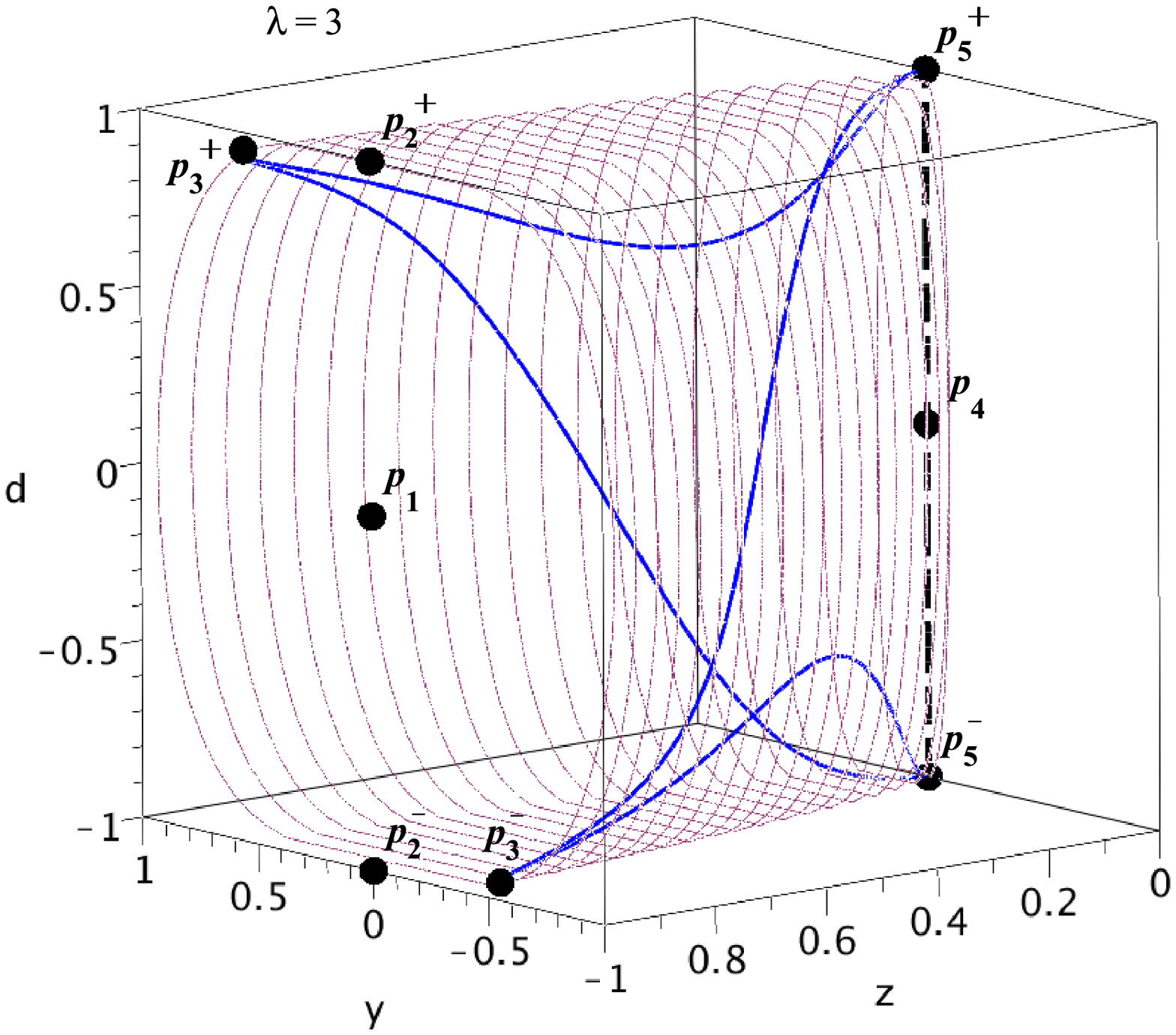}
\includegraphics[width=7cm]{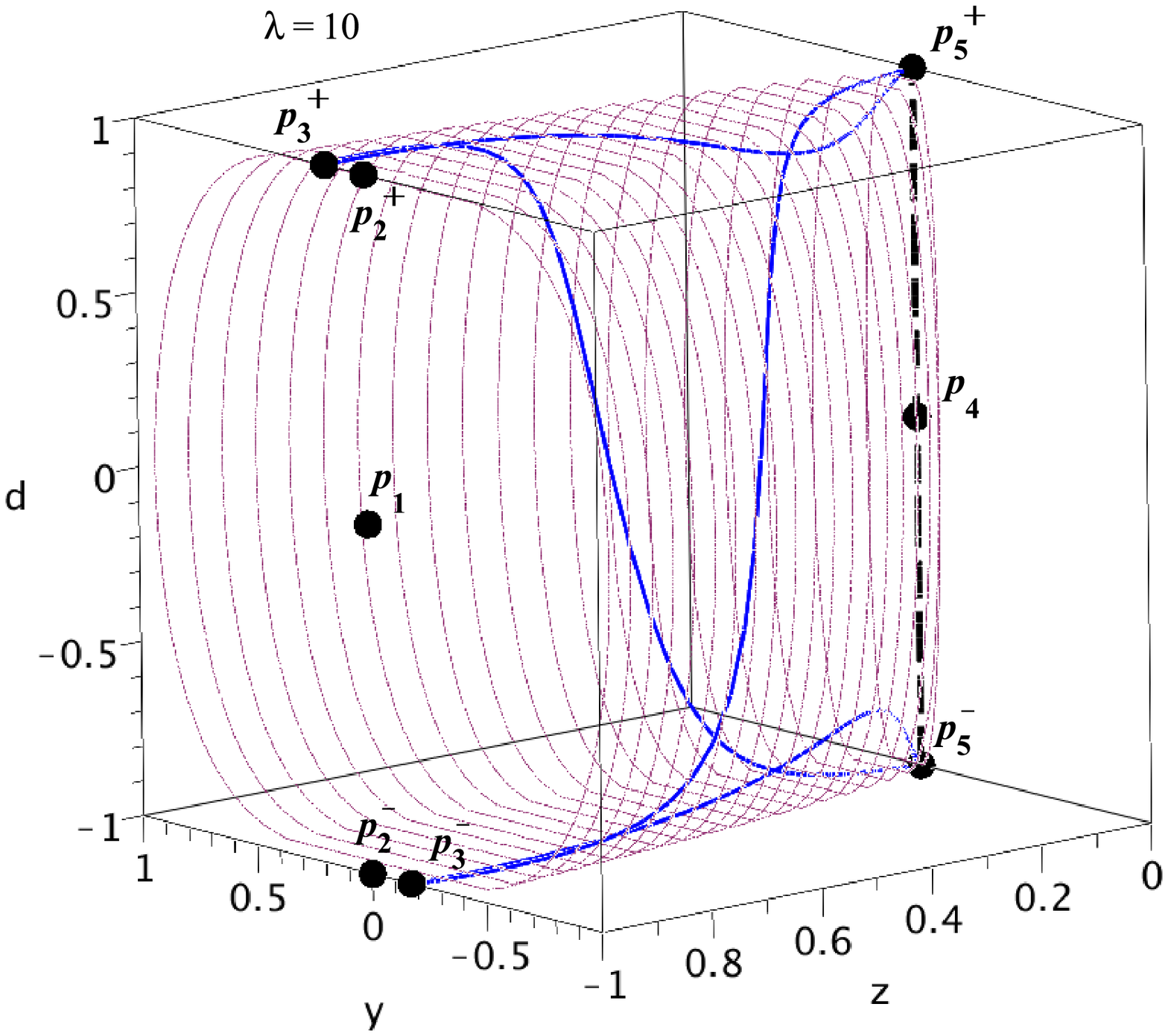}
\caption{The trajectories in the phase space of our model for the cases $\lambda=0$, $\lambda=1$, $\lambda=3$ and $\lambda=10$. The dashed black line and the dash-dotted red curve show the critical lines $L_1$ and $L_2$, respectively.}\label{fig5}
\end{figure*}

In FIG.\ref{fig5}, we have plotted some of the possible trajectories of our universe for a few values of $\lambda$. In all of the figures, the trajectories start from the unstable critical points $P_5^\pm$, but their final points are not unique. As it is obvious, for $\lambda=0$, the universe tends to reach the stable critical line $L_2$, or the stable critical points $P_6^\pm$. But as $\lambda$ starts to increase, $L_2$ and $P_6^\pm$, do not exist anymore. At these states, the trajectories tend to come to the attractors $P_3^\pm$, which coincide with $P_6^\pm$, at the limit $\lambda\rightarrow0$, and demonstrate a scalar field dominated solution. As we discussed in \ref{cp}, $P_3^\pm$, are potential dominated solutions for $\lambda\rightarrow0$, and kinetic dominated solutions for $\lambda\rightarrow\infty$, with $w_{tot}=-1+\lambda^2y_\ast^2$. Also, their positions in the phase volume depend on the value of $\lambda$. So, along with increasing $\lambda$, $P_3^\pm$, move in the plane $z=1$, until at the limit $\lambda\rightarrow\infty$, they approach $P_2^\pm$, while they are still attractors. If the universe evolves fast enough so that it comes to $P_3^\pm$, before $\lambda=\sqrt{2\sqrt3}$, or $w_{tot}=-1/3$, it certainly experiences a DE dominated era, but as soon as $w_{tot}$ crosses the line $-1/3$, or similarly $\lambda$, becomes greater than $\sqrt{2\sqrt3}$, another phase transition happens. The universe enters a decelerating expansion era, and stays at this phase forever. Presently, we know from observations such as the type Ia supernova \cite{Riess}, the cosmic microwave background radiation \cite{Spergel}, and so on, that we are in an accelerating expansion phase at the moment. Thus, according to our findings in this manuscript, we must wait for this phase transition in the future.

But, with attention to the model under consideration, there is another possible behavior for our universe. If it does not evolve fast enough, it never experiences a DE dominated era, because it reaches $P_3^\pm$, when $\lambda$, is greater than $\sqrt{2\sqrt3}$. Apparently, this is not the case that is happening for our universe, but it is one of the solutions of our model that may become important in future research.

\section{conclusion}\label{s4}

In this article, we have investigated a tachyonic RSII braneworld model in the context of the dynamical system approach. After introducing the model under consideration, we rewrote the main equations in terms of four new dimensionless variables $y$, $z$, $d$ and $\lambda$, which we introduced to set an autonomous system of ordinary differential equations. Then, we divided our discussion into two different situations $\lambda=constant$, and $\lambda=\lambda(\phi)$. In the first situation, that was related to an inverse square potential, we obtained eight critical points $P_1$, $P_2^\pm$, $P_3^\pm$, $P_4$, $P_5^\pm$, and one critical subset $L_1$, and calculated their eigenvalues. Then, we dealt with their cosmological interpretations and stability characteristics, both analytically and numerically.

In the second situation, we assumed a varying $\lambda$, and specifically we utilized a Gaussian tachyonic potential. We found two additional critical points $P_6^\pm$ and another critical line $L_2$, for the maximum of the potential or in other words for the case $\lambda=0$. We understood that they all are DE dominated attractors. But since $\lambda$ was considered to vary from zero to infinity, we saw that the fate of the universe changes instantly. We found that the trajectories want to reach $P_6^\pm$ or $L_2$, for $\lambda=0$, and since then along with increasing $\lambda$, they tend to come to $P_3^\pm$, which themselves depend on $\lambda$ and therefore move in the plane $z=1$.

Also, we found that the speed of the evolution of the universe has an important effect on its fate, so that if the trajectories get to the $P_3^\pm$, before $\lambda$ reaches the value $\sqrt{2\sqrt3}$, the universe experiences a DE dominated era and then enters a decelerating expansion phase. Otherwise, it never experiences an accelerating phase of expansion at all which apparently is inconsistent with observations.

\acknowledgments The authors thank Nelson J. Nunes for his valuable and helpful comments.

\nocite{*}
\bibliographystyle{spr-mp-nameyear-cnd}
%\bibliography{myref}
\bibliography{biblio-u1}

\begin{thebibliography}{}

\bibitem{Parsons} P. Parsons and J. D. Barrow, Phys. Rev. D 51, 6757 (1995).
\bibitem{Faraoni} V. Faraoni, Phys. Rev. D 53, 6813 (1996).
\bibitem{Steer} D. A. Steer and F. Vernizzi, Phys. Rev. D 70, 043527 (2004).
\bibitem{Emami} R. Emami, H. Firouzjahi, S. M. S. Movahed and M. Zarei, J. Cosmol. Astropart. Phys. 1102, 005 (2011).
\bibitem{Padilla} L. E. Padilla, J. A. Vazquez, T. Matos and G. German, J. Cosmol. Astropart. Phys. 05, 056 (2019).
\bibitem{Caldwell} R. R. Caldwell, R. Dave and R. J. Steinhardt, Phys. Rev. Lett. 80, 1582 (1998).
\bibitem{Caldwell2} R. R. Caldwell, Phys. Lett. B 545, 23 (2002).
\bibitem{Picon} C. Armendariz-Picon, V. Mukhanov and P. J. Steinhardt, Phys. Rev. D 63, 103510 (2001).
\bibitem{Padmanabhan} T. Padmanabhan, Phys. Rev. D 66, 021301 (2002).
\bibitem{Sen} A. Sen, Phys. Scripta. T 117, 70 (2005).
\bibitem{Feng} B. Feng, X. L. Wang and X. M. Zhang, Phys. Lett. B 607, 35 (2005).
\bibitem{Elizadle} E. Elizadle, S. Nojiri and S. D. Odintsov, Phys. Rev. D 70, 043539 (2004).
\bibitem{Kamenshchik} A. Kamenshchik, U. Moschella and V. Pasquier, Phys. Lett. B 511, 265 (2001).
\bibitem{Bento} M. C. Bento, O. Bertolami and A. A. Sen, Phys. Rev. D 66, 043507 (2002).
\bibitem{Cohen} A. G. Cohen, D. B. Kaplan and A. E. Nelson, Phys. Rev. Lett. 82, 4971 (1999).
\bibitem{Li} M. Li, Phys. Lett. B 603, 1 (2004).
\bibitem{Wei} H. Wei and R. G. Cai, Phys. Lett. B 663, 1 (2008).
\bibitem{Wei2} H. Wei and R. G. Cai, Phys. Lett. B 660, 113 (2008).
\bibitem{Gao} C. Gao, F. Wu, X. Chen and Y. G. Shen, Phys. Rev. D 79, 043511 (2009).
\bibitem{Gibbons} G.W. Gibbons, Phys. Lett. B 537, 1 (2002).
\bibitem{Bagla} J. S. Bagla, H. K. Jassal and T. Padmanabhan, Phys. Rev. D 67, 063504 (2003).
\bibitem{Das} A. Das, S. Gupta, T. D. Saini and S. Kar, Phys. Rev. D 72, 043528 (2005).
\bibitem{Farajollahi} H. Farajollahi, A. Ravanpak and G. F. Fadakar, Mod. Phys. Lett. A 26, 1125 (2011).
\bibitem{Farajollahi2} H. Farajollahi, A. Ravanpak and G. F. Fadakar, Astrophys. Space Sci. 336, 461 (2011).
\bibitem{Farajollahi3} H. Farajollahi, A. Salehi, F. Tayebi and A. Ravanpak, J. Cosmol. Astropart. Phys. 05, 017 (2011).
\bibitem{Farajollahi4} H. Farajollahi, A. Ravanpak and G. F. Fadakar, Phys. Lett. B 711, 225 (2012).
\bibitem{Ravanpak} A. Ravanpak and F. Salmeh, Phys. Rev. D 89, 063504 (2014).
\bibitem{Arkani} N. Arkani-Hamed, S. Dimopoulos and G. Dvali, Phys. Lett. B 429, 263 (1998).
\bibitem{Randall} L. Randall and R. Sundrum, Phys. Rev. Lett. 83, 4690 (1999).
\bibitem{Randall2} L. Randall and R. Sundrum, Phys. Rev. Lett. 83, 3370 (1999).
\bibitem{Setare} M. R. Setare, A. Ravanpak and H. Farajollahi, Gravit. Cosmol. 24, 52 (2018).
\bibitem{Farajollahi5} H. Farajollahi and A. Ravanpak, Astrophys. Space Sci. 349, 961 (2014).
\bibitem{Farajollahi6} H. Farajollahi, A. Ravanpak and G. F. Fadakar, Astrophys. Space Sci. 348, 253 (2013).
\bibitem{Farajollahi7} H. Farajollahi and A. Ravanpak, Phys. Rev. D 84, 084017 (2011).
\bibitem{Ping} Y. Ping, L. Xu, H. Liu and Y. Shao, Int. J. Mod. Phys. D 17, 11 (2008).
\bibitem{Campo} S. del Campo and R. Herrera, Phys. Lett. B 670, 266 (2009).
\bibitem{Barrow} J. D. Barrow, A. R. Liddle and C. Pahud, Phys. Rev. D 74, 127305 (2006).
\bibitem{Saridakis} E. N. Saridakis, Phys. Lett. B 660, 138 (2008).
\bibitem{Ravanpak2} A. Ravanpak, H. Farajollahi and G. F. Fadakar, Astrophys. Space Sci. 361, 43 (2016).
\bibitem{Saridakis2} E. N. Saridakis, J. Cosmol. Astropart. Phys. 04, 020 (2008).
\bibitem{Wainwright} J. Wainwright and G. F. R. Ellis, Dynamical Systems in Cosmology, Cambridge University Press, London 1997.
\bibitem{Coley} A. A. Coley, Dynamical Systems and Cosmology, vol. 291 of Astrophysics and Space Science Library, Springer Netherlands, Dordrecht 2003.
\bibitem{Biswas} S. K. Biswas and S. Chakraborty, Gen. Relativ. Gravit. 47, 22 (2015).
\bibitem{Zhang} K. Zhang, P. Wu and H. Yu, Phys. Lett. B 690, 229 (2010).
\bibitem{Zhang2} H. Zhang and Z. H. Zhu, Phys. Rev. D 75, 023510 (2007).
\bibitem{Nozari} K. Nozari, F. Rajabi and K. Asadi, Class. Quantum Grav. 29, 175002 (2012).
\bibitem{Ravanpak3} A. Ravanpak and G. F. Fadakar, Mod. Phys. Lett. A 34, 1950105 (2019).
\bibitem{Chimento} L. P. Chimento, R. Lazkoz, R. Maartens and I. Quiros, J. Cosmol. Astropart. Phys. 09, 004 (2006).
\bibitem{Copeland} E. J. Copeland, M. Sami and S. Tsujikawa, Int. J. Mod. Phys. D 15, 1753 (2006).
\bibitem{Copeland2} E. J. Copeland, M. R. Garousi, M. Sami and S. Tsujikawa, Phys. Rev. D 71, 043003 (2005).
\bibitem{Nozari} K. Nozari and N. Rshidi, Phys. Rev. D 88, 023519 (2013).
\bibitem{Gonzalez} T. Gonzalez, T. Matos, I. Quiros and A. Vazquez-Gonzalez, Phys. Lett. B 676, 161 (2009).
\bibitem{Quiros} I. Quiros, R. G. Salcedo, T. Matos and C. Moreno, Phys. Lett. B 670, 259 (2009).
\bibitem{Ravanpak4} A. Ravanpak and G. F. Fadakar, Class. Quantum Grav., 36, 235003 (2019).
\bibitem{Gumjudpai} B. Gumjudpai, Gen. Relativ. Gravit. 36, 747 (2004).
\bibitem{Mizuno} S. Mizuno, S. J. Lee and E. J. Copeland, Phys. Rev. D 70, 043525 (2004).
\bibitem{Bilic} N. Bilic, S. Domazet and G. Djordjevic, Class. Quant. Grav. 34, 165006 (2017).
\bibitem{Abramo} L. R. W. Abramo and F. Finelli, Phys. Lett. B 575, 165 (2003).
\bibitem{Riess} A. G. Riess et al., Astron. J. 116, 1009 (1998).
\bibitem{Spergel} D. N. Spergel et al., Astrophys. J. Suppl. Ser. 148, 175 (2003).

\end{thebibliography}

\end{document}